\documentclass[a4paper,11pt]{article}
\pdfoutput=1
\usepackage{color}

\usepackage{jcappub}

\providecommand{\e}[1]{\ensuremath{\times 10^{#1}}}
\newcommand{\sete}[1]{{\ensuremath{\mathcal{#1}_E}}}
\newcommand{\ivel}[1]{{\ensuremath{\dot{#1}_0}}}
\newcommand{\mpl}{{\ensuremath{M_\mathrm{Pl}}}}

\newcommand{\chaosprev}{Ramos:2001zw, Clesse:2009ur}

\newcommand{\eom}{\eqref{eqn:hybrideom}--\eqref{eqn:hybridv} }

\title{Initial conditions and sampling for multifield inflation}

\author{Richard Easther and}
\author{Layne C. Price}

\affiliation{Department of Physics \\  University of Auckland \\ Private Bag 92019 \\  Auckland, New Zealand}

\emailAdd{r.easther@auckland.ac.nz}
\emailAdd{lpri691@aucklanduni.ac.nz}


\begin{document}

\abstract{We investigate the initial conditions problem for multifield inflation.  In these scenarios the pre-inflationary dynamics can be chaotic,  increasing the sensitivity of the onset of inflation to the initial data even in the homogeneous limit.  To analyze physically equivalent scenarios we compare initial conditions at fixed energy.  This ensures that each trajectory is counted once and only once, since the energy density decreases monotonically.   We present a full analysis of hybrid inflation that reveals a greater degree of long range order in the set of ``successful'' initial conditions than was previously apparent.  In addition, we explore the effective smoothing scale for the fractal set of successful initial conditions induced by the finite duration of the pre-inflationary phase. The role of the prior information used to specify the initial data is discussed in terms of Bayesian sampling.
}

\maketitle
\flushbottom

\section{Introduction}
\label{sect:introduction}

The standard hot big bang is synonymous with the Friedmann-Lema{\^i}tre-Robertson-Walker (FLRW) metric.  This model imposes maximally symmetric initial conditions on the metric and the mass-energy distribution, as specified on an arbitrary initial spatial hypersurface.  These initial conditions  acausally correlate spacelike-separated regions and require further fine-tuning for the Universe to be spatially flat at late times, leading to the well-known horizon and flatness problems. Famously, these problems are resolved by inflation \cite{Starobinsky:1980te,Guth:1980zm,Linde:1981mu,Albrecht:1982wi}, which  grafts a phase of accelerated expansion onto the very early universe, setting the stage for the standard cosmology.  During inflation the comoving Hubble volume contracts and the visible universe is driven toward the spatially flat FLRW universe.

Given that inflation attempts to explain the otherwise {\em ad hoc} initial conditions of the standard hot big bang, a viable inflationary mechanism must itself be free of tunings.  Tunings can appear as  technically unnatural parameter values in the inflaton sector  or the need for a special pre-inflationary field configuration:  the latter question is the focus of this paper.  Inflationary models with unnatural initial conditions are at best incomplete and, at worst,  not viable as descriptions of the early universe.   Moreover, the level of tuning required to ensure the onset of inflation can differ substantially between scenarios with largely degenerate observational predictions,  providing a possible mechanism for discriminating between them.

The initial conditions problem arises even in the purely homogeneous limit.  For instance, chaotic inflation \cite{Linde:1983gd} begins for a large range of initial field values, but new inflation with the Coleman-Weinberg potential \cite{Linde:1981mu,Albrecht:1982wi} requires a special initial state.  Inflationary models with multiple scalar degrees of freedom introduce a further level of complexity.  With two or more fields the homogeneous dynamics are potentially  chaotic,  as first pointed out in Ref.~\cite{Easther:1997hm} and also discussed by Refs~\cite{\chaosprev}.  Chaos is synonymous with sensitive dependence on initial conditions, rendering multifield models qualitatively different from their single field counterparts. Multifield scenarios are widely studied and more natural in many settings.  In particular,  string theoretic inflationary scenarios often possess many scalar degrees of freedom.  Further, even if a  model has an effective single-field description once inflation is underway, the pre-inflationary phase may contain many interacting fields.   Several analyses of the  initial conditions problem for multifield inflation exist \cite{Lazarides:1996rk, Lazarides:1997vv, Tetradis:1997kp, Mendes:2000sq, Ramos:2001zw,Clesse:2008pf,Clesse:2009ur,Agarwal2011} and we return to this question here.

    \begin{table}
      \centering
      \begin{tabular}{| l  l | }
        \hline
        Set & Description \\
        \hline
        $\mathcal{I} \dots \dots .$             & Initial conditions surface, however defined \\
        $\mathcal{Z} \dots \dots .$            & Set of initial conditions with zero velocity \\
              $\sete{C} \dots \dots$            & Set of initial conditions with equal energy $E$ \\
        $\sete{S} \dots \dots$            & Successfully inflating subset of \sete{C} \\
              $\sete{F} \dots \dots$            & Non-inflating subset of \sete{C} \\
        $\sete{B} \dots \dots$            & Boundary between \sete{S} and \sete{F} \\
        \hline
      \end{tabular}
      \caption{Subsets of phase space: notation.}
    \label{table:legend}
  \end{table}

We need to sample the ``initial conditions space'' $\mathcal{I}$ for these scenarios, determining  the overall fraction that inflates and the topology of the inflationary region within this space.   A homogeneous, spatially flat universe  containing $N$ scalar fields $\phi_i$ with arbitrary interactions has  $2N$ independent degrees of freedom since the scale factor can be eliminated by the 0-0 Einstein equation.  The solutions to the equations of motion --- called trajectories or orbits --- are non-intersecting curves that fill the $2N$--dimensional phase space.  Yet, the initial field values and velocities are not independent or identically distributed (iid) random variables as different points in $\mathcal I$ are correlated by the solutions to the field-equations, \emph{i.e.}, many points belong to the same trajectory. (See Table~\ref{table:legend} for a summary of our notation.) 

The phase space is foliated by surfaces of equal energy, \sete{C}. The energy density $\rho=E^4$ of FLRW universes is monotonic, decreasing in a homogeneous universe as
\begin{equation}
	\label{eqn:rhodot}
  \dot{\rho}(t) = -3H \sum_{i=1}^N \dot \phi_i^2,
  \end{equation}
  where the Hubble parameter $H \propto E^2$ and overdots denote derivatives with respect to coordinate time $t$.  For a specific energy $E$, orbits intersect $\sete{C}$ once and only once, identifying each  point on  $\sete{C}$ with a unique solution to the equations of motion.  To build a well-defined sample of trajectories we choose initial conditions from the constraint surface $\sete{C}$.

Many previous treatments of the multifield initial conditions problem \cite{Lazarides:1996rk,Lazarides:1997vv,Tetradis:1997kp,Mendes:2000sq,Clesse:2008pf} have been based on $\mathcal{Z}$, the $N$--dimensional  subset of $\mathcal{I}$ on which all velocities vanish simultaneously. Although the set of inflationary trajectories  intersecting $\mathcal{Z}$  is easier to sample than \sete{C}, orbits for which all velocities vanish at the same instant are not generic, given the finite duration of the pre-inflationary era.  Many orbits thus never intersect $\mathcal{Z}$, while in principle others may intersect it multiple times, correlating apparently distinct points --- issues that cannot arise when sampling from \sete{C}.  
By contrast, Ref.~\cite{Clesse:2009ur} samples the full phase space $\mathcal{I}$ and also varies the parameters in the potential itself, effectively marginalizing over the energy scale $E$.   The current paper is the first analysis of the initial conditions problem for multifield inflation that does not either (a) study a lower-dimensional surface in the initial conditions space (however defined) or (b) sample the entirety of $\mathcal{I}$, disregarding the fact that different points  belong to the same solution of the field equations.\footnote{Tetradis~\cite{Tetradis:1997kp} presents a single equal energy slice, although a degree of freedom was removed by requiring the velocities to be equal, similar to the projections we introduce for convenience in Figs~\ref{fig:slice1e-2}--\ref{fig:slice1e-5}.  }    

Any description of the primordial universe  breaks down above some energy scale, and this scale defines the appropriate initial conditions hypersurface \sete{C} for a well-specified model.\footnote{If the potential has one or more local minima where $V_\Lambda>0$, choosing $E^4 < V_\Lambda$  will necessarily exclude all trajectories which evolve toward these minima.}     This energy will be associated with some scale in the particle physics sector, such as the characteristic size of extra dimensions, the string scale,  next-to-leading order corrections to Einstein gravity, or ultimately its breakdown at the Planck scale. Points on $\sete{C}$ are thus physically commensurate, whereas  points in $\mathcal{I}$ span several orders of magnitude in energy.    Qualitatively, we will also  find that the set of successfully inflating points has a simpler structure and more obvious long range order when chosen from  \sete{C} rather than $\mathcal{Z}$, allowing us to better understand the underlying cosmological dynamics.   

Beyond the choice of initial conditions surface,  we must also specify the prior probability distributions (in Bayesian terms) for the initial field values and velocities.  If $\omega$ is a probability distribution that weights an initial condition $\mathbf{x}_0$ according to how well its final state matches the observed universe, then the expected value of $\omega$ over initial conditions $\mathbf{x}_0 \in \sete{C}$ is
\begin{equation}
  \left\langle \omega (\sete{C}) \right\rangle = \int_{\sete{C}} \omega (\mathbf{x}_0) P_E(\mathbf x_0) d^{N} x_0 \approx \frac{1}{n} \sum_{i=1}^n \left\{ \omega (\mathbf{x}_0^{(i)}) \right\}_{\mathbf{x}_0^{(i)} \in C_E},
  \label{eqn:expectation}
\end{equation}
where $P_E$ is the prior probability distribution for initial conditions on the constraint surface $\sete{C}$ and the sum is evaluated at $n$ points sampled from $C_E$.  The prior in Eq.~\eqref{eqn:expectation} acts as the probability density function for initial conditions on the space of FLRW universes. The form of $P_E$ is only weakly constrained by fundamental considerations.  The freedom to choose $P_E$ is analogous to the measure problem in the multiverse~\cite{Gibbons1987,Hawking:1987bi,Gibbons:2006pa,Freivogel:2011eg,Schiffrin:2012zf}, albeit restricted to the subspace of homogeneous FLRW universes.   In many previous works the prior is often not directly discussed, and  thus implicitly defined as a uniform distribution on the   initial conditions. We consider several possible choices of prior (all of which are uninformative) and vary the energy of the surfaces $\sete{C}$.  We find that the choice of prior significantly alters the fraction of trajectories that lead to inflation, potentially distorting  conclusions about the extent to which a given inflationary model requires fine-tuned initial conditions.

  In what follows we work with a widely studied two-field model: canonical hybrid or false-vacuum inflation \cite{Kofman:1986wm,Linde:1993cn,Copeland:1994vg}.  We relate the initial conditions problem to that of determining the (fractal) topology and the geometry of the subset of points $\sete{S} \subset \sete{C}$ that successfully inflate, since this is independent of the choice of prior.   Like Refs~\cite{Clesse:2008pf,Clesse:2009ur} we see that $\sete{S}$ has a fractal topology due to the presence of chaos in the underlying dynamical system,  demonstrating that hybrid inflation has regions of phase space where orbits are highly sensitive to their initial conditions and confirming the results of Ref.~\cite{Easther:1997hm}.   Hybrid inflation is associated with a blue power spectrum\footnote{Although see Refs~\cite{Clesse:2010iz,Kodama:2011vs}.}  at odds with recent astrophysical data \cite{Story:2012wx,Hinshaw:2012fq,Bennett:2012fp,Sievers:2013wk,Ade:2013xsa}.   However, our primary focus is not  hybrid inflation itself, but  developing tools that can be used to understand the initial conditions problems in generic models of multifield inflation. We use this model because (a) it is the prototypical multifield model with chaotic dynamics and a narrowly defined inflationary attractor; (b) we are primarily interested in the onset of inflation; and (c) to make contact with previous work.  
  
  The pre-inflationary universe is dissipative, so the fractal structure must have a nontrivial scale dependence: there is necessarily a minimum scale below which  two nearby trajectories will remain correlated until they reach either the inflationary attractor or a minimum of the potential, smoothing $\sete{S}$ below this scale.    Conversely, while we assume classical homogeneity, quantum fluctuations    prevent the universe from being {\em perfectly\/} smooth.  If $\sete{S}$ has structure on scales smaller than a typical fluctuation we cannot sensibly define the homogeneous limit for this system. Consequently, we propose a sampling technique that identifies regions where $\sete{S}$ has structure below this minimum scale.

The paper is arranged as follows: in Section~\ref{sect:dynamics} we review hybrid inflation and discuss its dynamics.  In Section~\ref{sect:results} we describe our numerical methods, characterize the properties of the set of inflationary trajectories with different energies and priors $P_E$, and investigate the fractal  dimension of \sete{S}.  In Section~\ref{sect:conclusion} we discuss the implication of our results and identify  future lines of enquiry.


\section{Inflationary dynamics}
\label{sect:dynamics}

For simplicity we  consider two homogeneous scalar fields, $\psi$ and the inflaton $\phi$, interacting through a potential $V(\psi,\phi)$ in a homogeneous FLRW universe.  The equations of motion are
\begin{equation}
	\label{eqn:hybrideom}
  \ddot{\phi} + 3H\dot{\phi} + \frac{\partial V}{\partial \phi} = 0  \qquad \text{and} \qquad
  \ddot{\psi} + 3H\dot{\psi} + \frac{\partial V}{\partial \psi} = 0,
\end{equation}
and the Hubble parameter $H$ can be eliminated by the 0-0 Einstein equation
\begin{equation}
  H^2 = \frac{8\pi}{3 M_\mathrm{Pl}^2} \left[ \frac{1}{2}\dot{\phi}^2 + \frac{1}{2}\dot{\psi}^2
  + V(\psi,\phi) \right],
  \label{eqn:einstein2}
\end{equation}
where $\mpl$ is the Planck mass.  Following Refs~\cite{Lazarides:1996rk, Lazarides:1997vv, Easther:1997hm,Tetradis:1997kp, Mendes:2000sq, Ramos:2001zw,Clesse:2008pf,Clesse:2009ur} we consider hybrid inflation \cite{Kofman:1986wm,Linde:1993cn,Copeland:1994vg} with the potential
\begin{equation}
	\label{eqn:hybridv}
  V(\psi,\phi) = \Lambda^4 \left[\left(1-\frac{\psi^2}{M^2} \right)^2 + \frac{\phi^2}{\mu^2}
  + \frac{\phi^2 \psi^2}{\nu^4} \right],
\end{equation}
with real parameters $\Lambda$, $M$, $\mu$, and $\nu$.  Inflation occurs in the ``inflationary valley" with $\psi \approx 0$ and  $|\phi| > \phi_c$, where  $\phi_c= \sqrt{2} \, \nu^2/M$ is the critical point at which the effective mass of $\psi$ becomes complex and inflation comes to an end.    The potential is symmetric under $\phi\rightarrow-\phi$ and $\psi\rightarrow-\psi$,  with two equivalent valleys for $\phi>\phi_c$ and $\phi<-\phi_c$ and minima at $\{\psi,\phi\}=\{\pm M,0\}$.  Orbits will either enter one of the false-vacuum inflationary valleys or evolve directly toward one of the true vacua.

We set the amplitude $A_s$ of the dimensionless power spectrum $\mathcal{P}_\mathcal{R}$ to be roughly compatible with the WMAP9 data \cite{Bennett:2012fp,Hinshaw:2012fq}, which fixes the potential energy scale.  This results in
\begin{equation}
  \label{eqn:powerspect}
  A_s \approx \frac{1}{24\pi^2 M_\mathrm{Pl}^4} \left( \frac{V}{\epsilon_V} \right) = \left( 2.43 \pm 0.08 \right) \e{-9},
\end{equation}
where $\epsilon_V = (M_\mathrm{Pl}^2/2)(V_{,\phi}/V)^2$ is the slow-roll parameter.
Setting $M=.03  \, \mpl$, $\mu=500  \, \mpl$, and $\nu=.015 \, \mpl$ and assuming perturbations are generated when $\psi \approx 0$ and $\phi \approx \phi_c$, we derive $\Lambda \approx 6.8\e{-6} \, \mpl$.\footnote{The super-Planckian value of $\mu$ is an artifact of this definition of the potential; the actual mass term is $m_\phi^2 = 2\Lambda^4/\mu^2 \approx 10^{-26}\, M_\mathrm{Pl}^2$, and safely sub-Planckian.}  Lastly, Ref.~\cite{Martin:2011ib} determined that quantum fluctuations dominate the classical field evolution in the inflationary valley when
\begin{equation}
  \label{eqn:lambdaquant}
  \Lambda > \Lambda_q \equiv 4 \pi \sqrt{3} M_\mathrm{Pl}^3 \frac{\phi_c}{\mu^2}.
\end{equation}
For our parameters $\Lambda_q = 9.6\e{-4} \, \mpl$, so the classical equations are self-consistent.

    \begin{figure}
      \centering
      \includegraphics{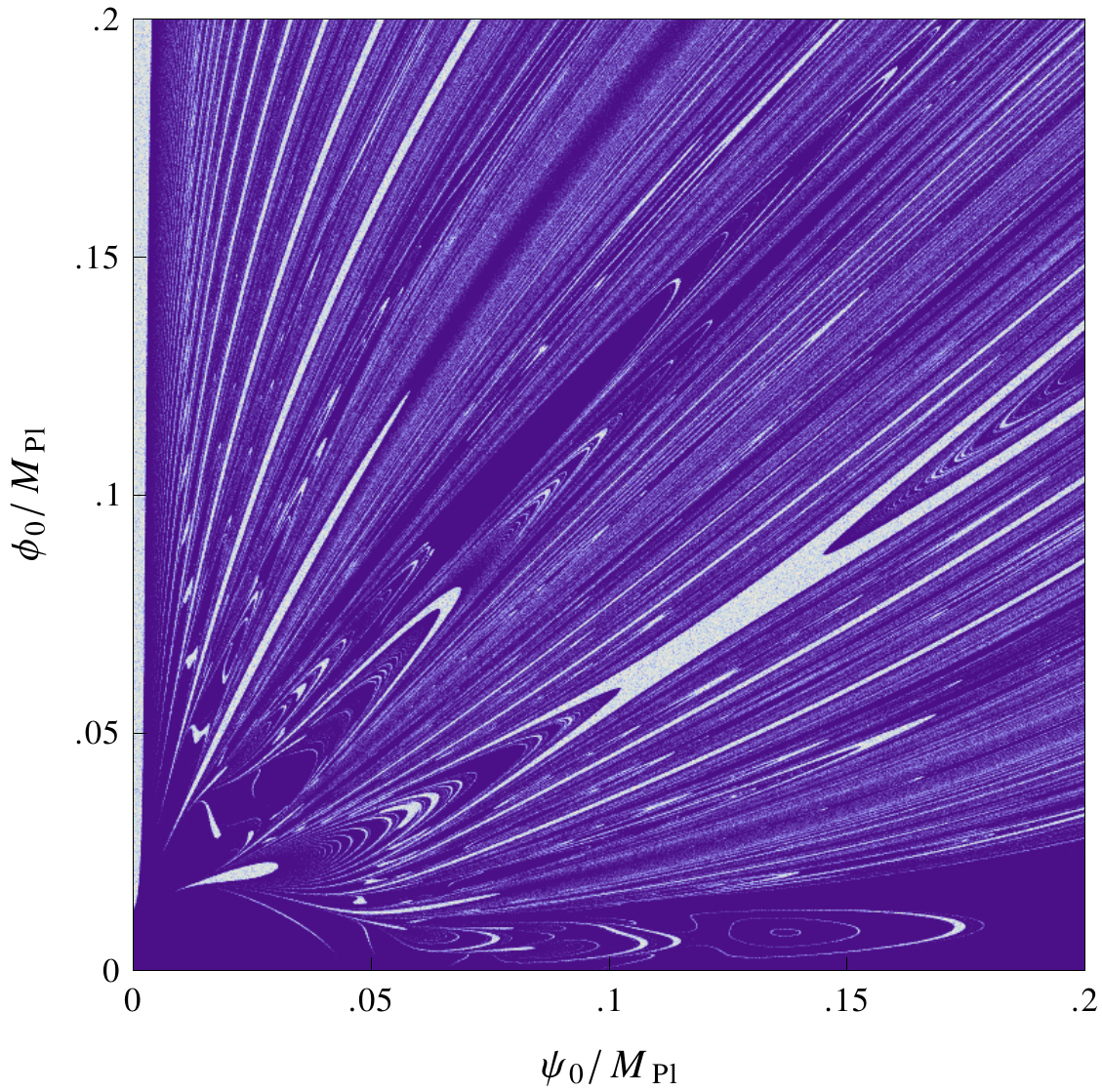}
      \caption{Distribution of successfully inflating initial conditions drawn from the zero-velocity slice $\mathcal{Z}$.  White areas have the highest number of successful points; darker regions have the fewest. This plot matches Fig.~(1) of Ref.~\cite{Clesse:2009ur}, with $M=.03\, \mpl$, $\mu=636\,\mpl$, and $\nu=.0173\,\mpl$.  The figures are similar (verifying our codes and algorithms) but are not expected to be identical, due to different binning procedures.}
      \label{fig:clessefig1}
    \end{figure}
The inflationary valley is a small subset of the total  phase space, which might suggest the model has a fine-tuning problem.  References \cite{Tetradis:1997kp,Lazarides:1996rk,Mendes:2000sq} considered sub-Planckian initial field values on the zero-velocity surface  $\mathcal{Z}$, pessimistically concluding that --- in the absence of effects that increase the friction experienced by the fields --- only trajectories which start inside the inflationary valley yield 60 e-folds of inflation.  By contrast, Ref.~\cite{Lazarides:1997vv} was more optimistic, showing that a supergravity-inspired hybrid inflation model has a significant number of ``successful'' points outside  the inflationary valley.  With more exhaustive sampling of $\mathcal{Z}$, subsequent studies by Clesse, Ringeval, and Rocher \cite{Clesse:2008pf,Clesse:2009ur} extended this optimistic conclusion to the potential~\eqref{eqn:hybridv}.  They showed that successful initial conditions are distributed in an intricate series of patches and fine lines outside the inflationary valley, with a fractal boundary separating inflating and non-inflating initial conditions.
The distribution of successfully inflating initial conditions on $\mathcal{Z}$, for a specific scenario from Ref. \cite{Clesse:2009ur}, is reproduced in Fig.~\ref{fig:clessefig1}.  The fine-tuning problem may also be less serious if the initial field values are assumed to be super-Planckian or if the interaction term dominates \cite{Felder:1999pv,Clesse:2008pf}.

References~\cite{Clesse:2009ur,Clesse:2010iz} also present a Markov Chain Monte Carlo (MCMC) sampling of all possible parameter choices and  sub-Planckian field configurations, including those with initial velocities.  The conclusion was that 60 e-folds of inflation is generic for the potential in Eq.~\eqref{eqn:hybridv} and fine-tuned initial conditions in the inflationary valley are not required.  Although we have argued that sampling from any two-dimensional subspace, such as $\mathcal Z$, is of limited benefit, sampling the whole four-dimensional space $\mathcal I$ may not be strictly necessary, even though an MCMC technique marginalizes the unknown initial energy $E$.  We instead choose to explore how fine-tuned the initial conditions must be when sampling from constraint surfaces $\sete{C}$ that incorporate the energy constraint Eq.~\eqref{eqn:rhodot}.



\section{Numerical results}
	\label{sect:results}

\subsection{Method}
\label{ssect:method}

We numerically integrate Eqs~\eom using a backward-difference formula implemented by the \textsc{Fcvode} package from the \textsc{Sundials} computing suite \cite{Hindmarsh:2005}.   We sample initial conditions from constraint surfaces \sete{C} with constant energy density
\begin{equation}
  \rho = \frac{1}{2} \ivel \psi^2 + \frac{1}{2} \ivel \phi^2 + V(\psi_0,\phi_0)=E^4,
  \label{eqn:energyeqn}
\end{equation}
where
\begin{equation}
  \label{eqn:chosenenergies}
  E = 10^{i} \, \mpl \qquad \mathrm{for} \qquad i \in \{-5,\dots,0\}.
\end{equation}
The last 60 e-folds of inflation occur at $\{\psi,\phi\} \approx \{0,\phi_c\}$ with $E \sim 10^{-6} \, \mpl$.  With $E= 10^{0} \mpl$ we are at the limit of classical Einstein gravity; we only include this case to illustrate the underlying dynamical system.

We stop integrating when  either (a) the orbit achieves more than 60 e-folds during inflation or (b) $\rho < \Lambda^4$ and the trajectory is trapped by the potential wells at $\{\psi,\phi\} = \{\pm M, 0 \}$.  Initial conditions which lead to 60 e-folds of inflation are ``successful'' and define the subset \sete{S}, while its complement --- the ``failed'' points --- comprise the subset  \sete{F}.\footnote{Again, additional constraints can be added, e.g. data matching for $n_s$, $r$, or other observables.} The boundary between these sets, whose properties determine the extent to which they ``mix,''  is denoted \sete{B}.

We select points randomly on the constraint surface as follows. We first draw $\phi_0$ and $\psi_0$ from the uniform distribution over $0 \le \{\psi_0,\phi_0\} \le .2 \, \mpl$, excluding any choices with $V(\phi_0,\psi_0) >E^4$.  The symmetry of the potential~\eqref{eqn:hybridv} allows the restriction to positive field values, whereas the upper bound is set to be consistent with Ref.~\cite{Clesse:2009ur}.  Given these initial field values, the kinetic energy is typically dominant unless $E \approx \Lambda \sim 10^{-6} \; \mpl$.  We apportion the remaining energy by drawing one of $v_1 \in \{\ivel{\phi}, \ivel{\psi}\}$ from a uniform prior on the range
\begin{equation}
  -\sqrt{2(E^4 -V_0)} \le v_1 \le \sqrt{2(E^4 -V_0)}
  \label{eqn:velrange}
\end{equation}
and giving the leftover energy to the other field velocity $v_2$ with the overall sign again chosen randomly.  We maintain the symmetry between $\ivel\phi$ and $\ivel\psi$ by alternating the order in which these velocity terms are set.\footnote{This prior generates the high tails in the velocity distributions seen in Fig.~\ref{fig:histogram}.  If the first velocity chosen is $v_1$, the second will be $v_2 =\pm \sqrt{2(E^4 -V_0) - v_1^2}$.  Since $v_1$ is uniformly distributed, $v_2$ is a quadratic distribution, favoring higher values.}  This procedure implicitly defines the initial priors $P_\mathrm{orig}$ on the constraint surfaces \sete{C}.

To estimate the size of inhomogeneous fluctuations at the initial energy $E$, we note that $\delta \phi \sim H/2\pi$ and $H \sim E^2/\mpl$ for a massless field in de Sitter space.  Similarly, the minimal variation in velocities is expected to be of order $H^2$ across a Hubble volume \cite{Martin:2011ib,Finelli:2010sh,Finelli:2008zg}. The fields $\psi$ and $\phi$ are not massless and the pre-inflationary universe is not de Sitter, but we can use this relationship to put an approximate lower bound on the homogeneity of the primordial universe.  In regions of \sete{S} whose typical scale in any phase space dimension is less than
\begin{equation}
  \Delta \equiv \{ \delta \psi_0, \delta \phi_0,\delta \ivel{\psi},\delta \ivel{\phi} \} = \frac{1}{2 \pi} \{H,H,H^2,H^2\}
  \label{eqn:Delta}
\end{equation}
the homogeneous approximation breaks down and further analysis is invalid or ambiguous.  Physically, in these regions we cannot self-consistently assume that the primordial universe is homogeneous. 



We exclude these regions from \sete{S} by sampling \sete{C} in clusters.  We first choose points from \sete{C} and integrate Eqs~\eqref{eqn:hybrideom}--\eqref{eqn:hybridv}.  At each point that successfully inflates we randomly draw 100 points within $\Delta$ of that point.\footnote{For the lowest values of $E$, $\left\{\delta \psi_0/\psi_{\mathrm{max}}, \, \delta \phi_0/\phi_{\mathrm{max}}, \, \delta \dot{\psi}_0/\dot{\psi}_{\mathrm{max}}, \, \delta \dot{\phi}_0/\dot{\phi}_{\mathrm{max}} \right\} \sim 10^{-10}$, which is far below the resolution of our figures.  We confirmed the accuracy of the \textsc{Fcvode} integrator in this domain by using an arbitrary precision integrator from Mathematica.} If any of these new points do not inflate, we conclude that the original point was a ``false'' (or perhaps ambiguous) positive.  Applying this simple stability check at various other points along the trajectory's evolution is straightforward, but computationally expensive.  Furthermore, the largest fluctuations occur at the highest energies, so testing the initial energy surface captures the most relevant effects.  Although this approach incorporates points lying near (but not actually on) our designated equal energy surface, we do not weight our conclusions by these secondary points.

\begin{figure}
     \includegraphics[width=.35\textwidth]{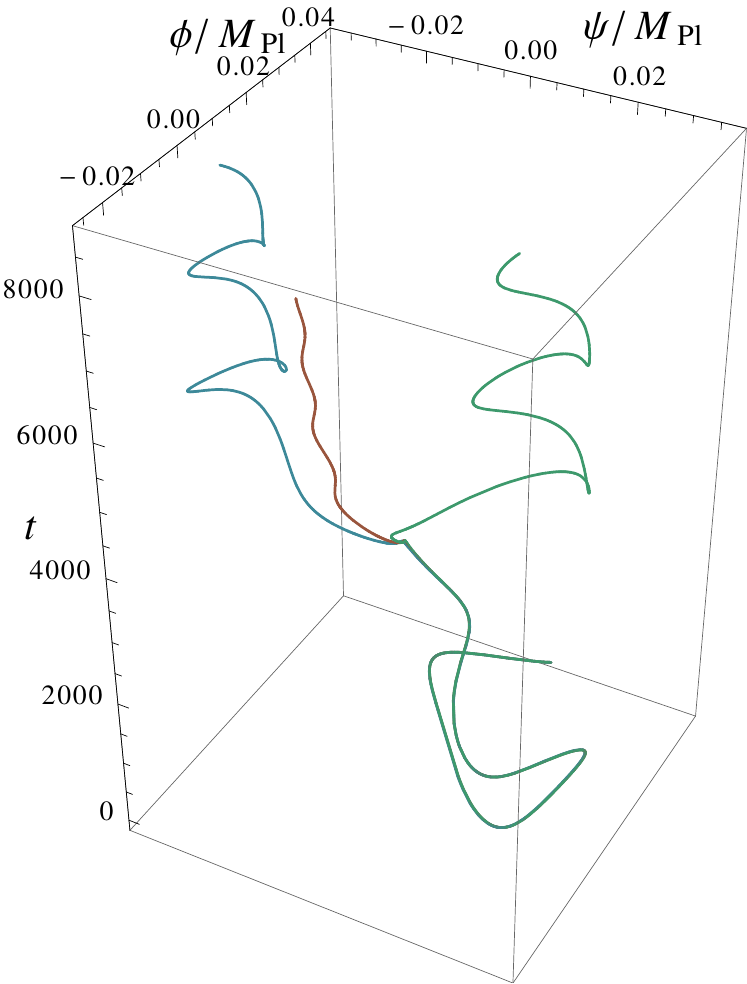}
     \includegraphics{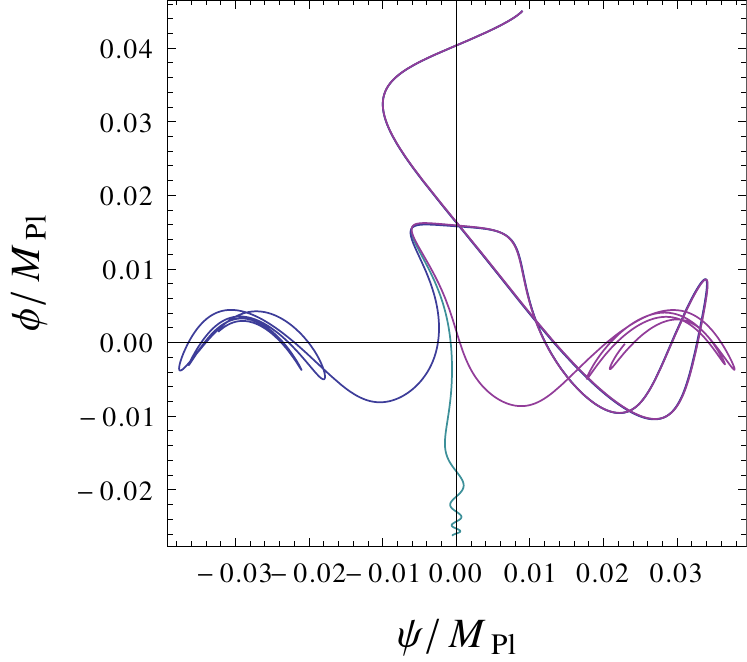}
     \caption{Parametric trajectory plots for three orbits at energy $E=10^{-5} \, \mpl$ initially separated by $10^{-8}\mpl$ in the field values and $10^{-8}M_\mathrm{Pl}^2$ in the field velocities.  These orbits are exponentially diverging, with each of the three trajectories branching at $t \approx 3600 \, M_\mathrm{Pl}^{-1}$; one goes to each of the global minima at $\psi=\pm M$ and one inflates.}
      \label{fig:exponentialdiv_small}
    \end{figure}

This analysis does not address the inhomogeneous initial conditions problem; it simply limits the extent to which the initial conditions can be self-consistently fine-tuned in a homogeneous universe, given that the chaotic dynamics of the potential may cause closely correlated trajectories to diverge exponentially.  Fig.~\ref{fig:exponentialdiv_small} shows three solutions of Eqs~\eqref{eqn:hybrideom}--\eqref{eqn:hybridv}, at $E=10^{-5} \, \mpl$ with initial field values which differ by only $10^{-8} \mpl$.  They eventually diverge, with each trajectory reaching a distinct end-state.  If an inflation model has a fractal \sete{S} or \sete{B} that is distributed in a complex manner over \sete{C}, then almost all successfully inflating initial conditions may be within $\Delta$ of an initial condition which does not inflate.


\subsection{Successful inflationary trajectories}
\label{ssect:trajs}

    \begin{table}
      \centering
      \begin{tabular}{| c  c  c  c  c  c |}
        \hline
        $E$ [$\mpl$] & $n_{\mathrm{succ}}$ [M] & $n_{\mathrm{total}}$ [M] & $n_{\mathrm{false}}$ [k] & $f_{\mathrm{true}}$ & $f_\mathrm{total}$  \\
        \hline
        $10^{0}$             & 0.00 & 2.01 & 1000.0 & 0.000  & 0.498 \\
        $10^{-1}$            & 1.00 & 2.24 & 72.7   & 0.447  & 0.479 \\
        $10^{-2}$            & 1.00 & 2.57 & 114.6  & 0.389  & 0.434 \\
        $10^{-3}$            & 1.00 & 3.34 & 105.9  & 0.300  & 0.331 \\
        $10^{-4}$            & 1.00 & 4.89 & 30.4   & 0.205  & 0.211 \\
        $\dagger \, 10^{-5}$ & 1.00 & 4.45 & 0.012  & 0.225  & 0.225 \\
        \hline
      \end{tabular}
      \caption{Total fraction of successfully inflating points sampled from priors $P_\mathrm{orig}$ on the equal energy slices \sete{C} --- both excluding ($f_{\mathrm{true}}$) and including ($f_\mathrm{total}$) false positives from \sete{S}.  Also shown are the number of successful points $n_{\mathrm{succ}}$, the number of false positives $n_{\mathrm{false}}$, and the combined number of fail points, false positives, and successful points $n_{\mathrm{total}}$.  The numbers $n_\text{succ}$ and $n_\text{total}$ are measured in millions [M] of points, $n_\text{false}$ is measured in thousands [k] of points, and the energy $E$ is in units of the Planck mass $\mpl$.  ($\dagger$) The sampling procedure deviates from an ``equal-area'' sample as $E \to \Lambda$.}
    \label{table:ratio}
  \end{table}

    \begin{figure}
     \centering
      \includegraphics{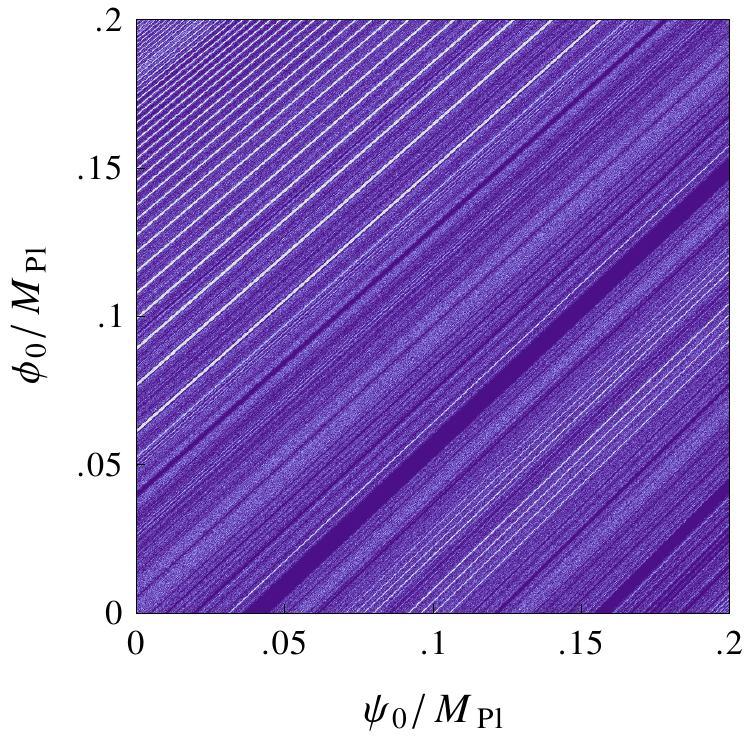}
      \includegraphics{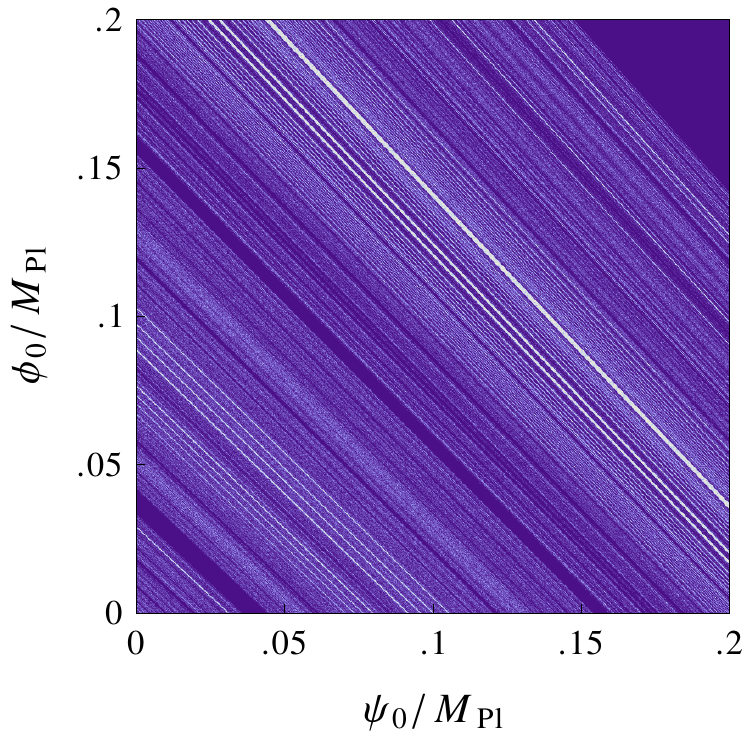}
      \includegraphics{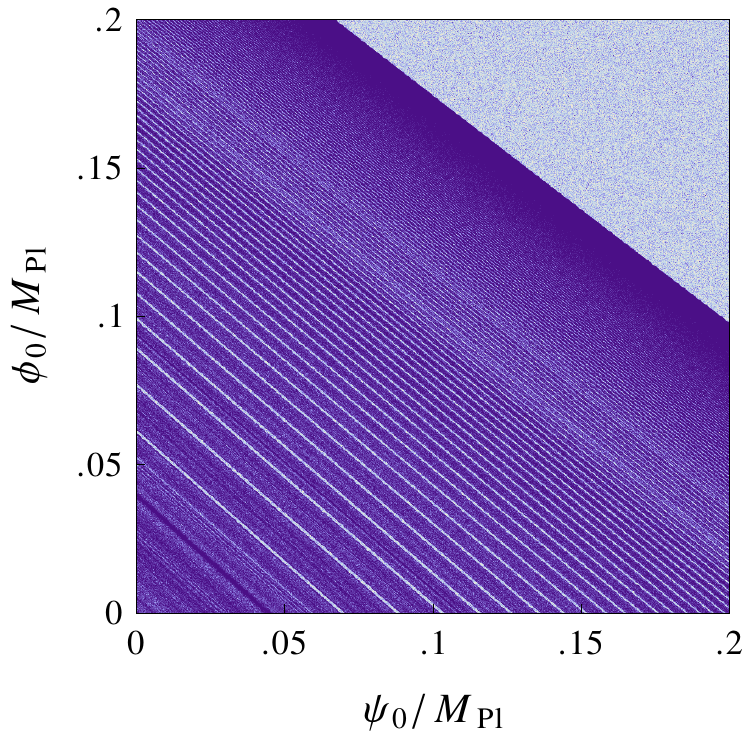}
      \includegraphics{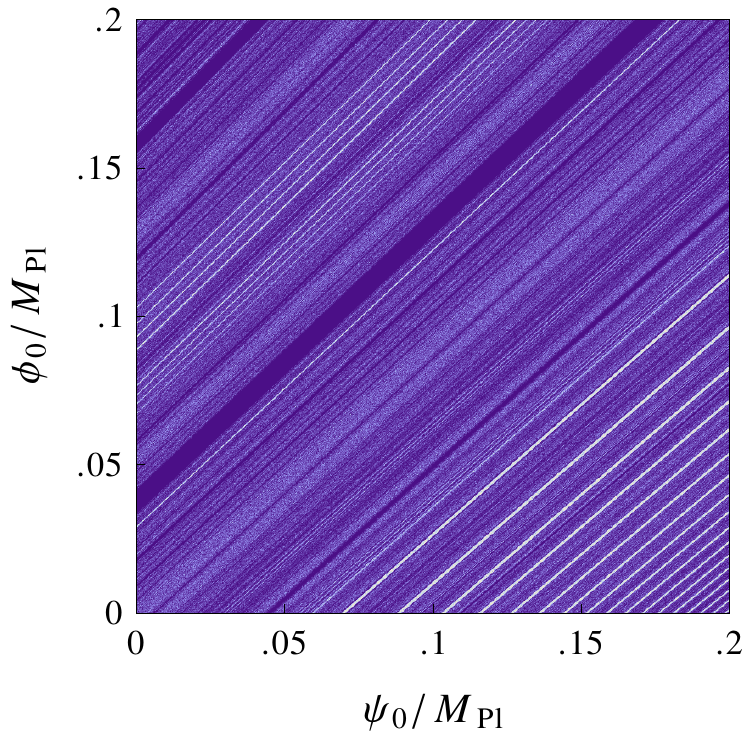}
      \caption{Two dimensional slicings of \sete{C} for $E= 10^{-2} \, \mpl$, including the ambiguous, ``false positive'' points in \sete{S}.  Parameters are $\Lambda = 6.8\e{-6} \, \mpl$, $M=.03  \, \mpl$, $\mu=500  \, \mpl$, and $\nu=.015 \, \mpl$.  The light and dark areas are regions that have a higher and lower density of points in \sete{S}, respectively.  The results have been binned over a $1000\times1000$ grid.  All velocities are of equal magnitude, however the left column has $\dot{\phi}>0$, the right column is at $\dot{\phi}<0$, the top row has $\dot{\psi}>0$, and the bottom row has $\dot{\psi}<0$.}
      \label{fig:slice1e-2}
    \end{figure}

    \begin{figure}
     \centering
      \includegraphics{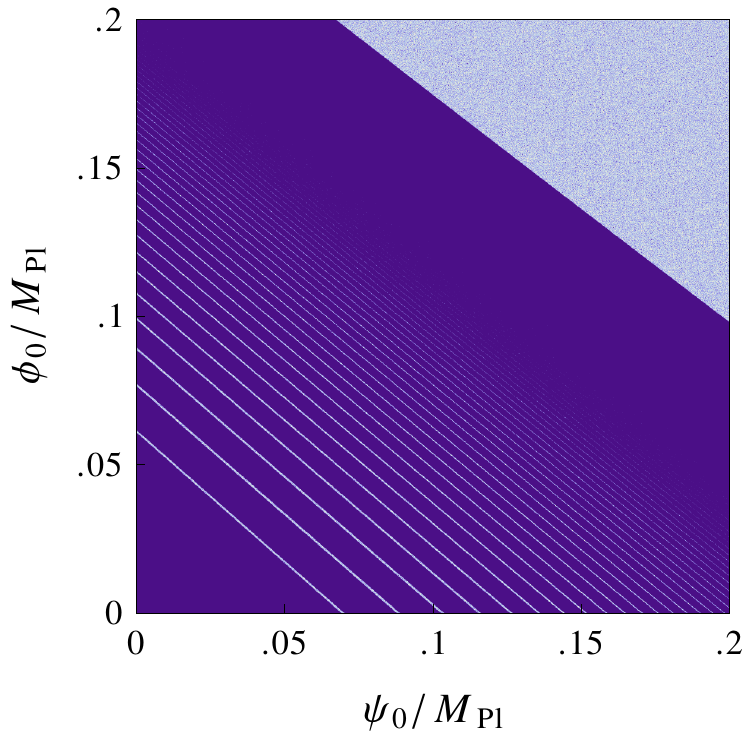}
      \includegraphics{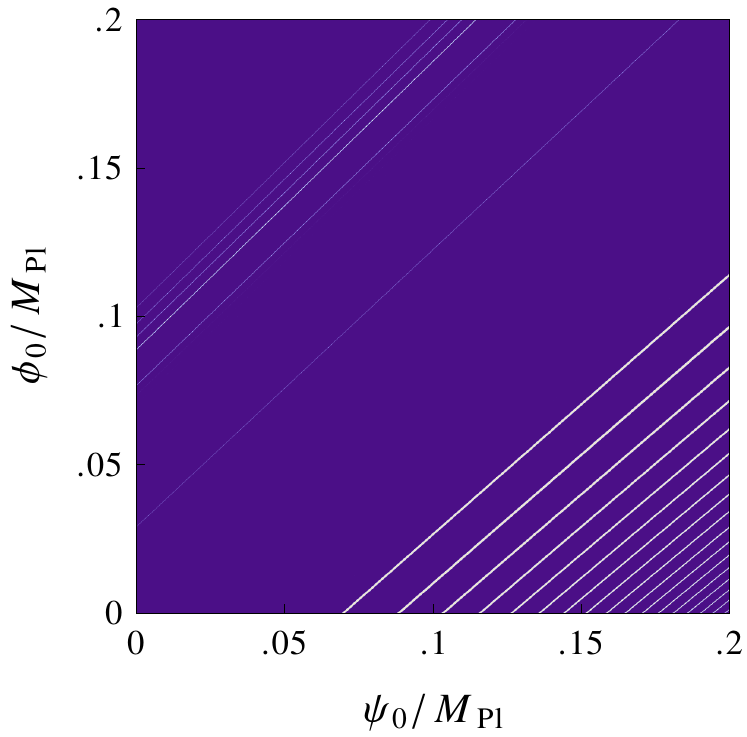}
      \caption{Two dimensional slicings of \sete{C} for $E= 10^{-2} \, \mpl$, excluding any ambiguous or ``false'' positives from the set of successfully inflating initial conditions, \sete{S}.  All velocities are of equal magnitude; the left panel has $\ivel{\psi}<0$ and $\ivel{\phi}>0$; and the right panel has $\{\ivel{\psi},\ivel{\phi}\}<0$.}
      \label{fig:slice1e-2_nofalsepos} 
    \end{figure}

    \begin{figure}
     \centering
      \includegraphics{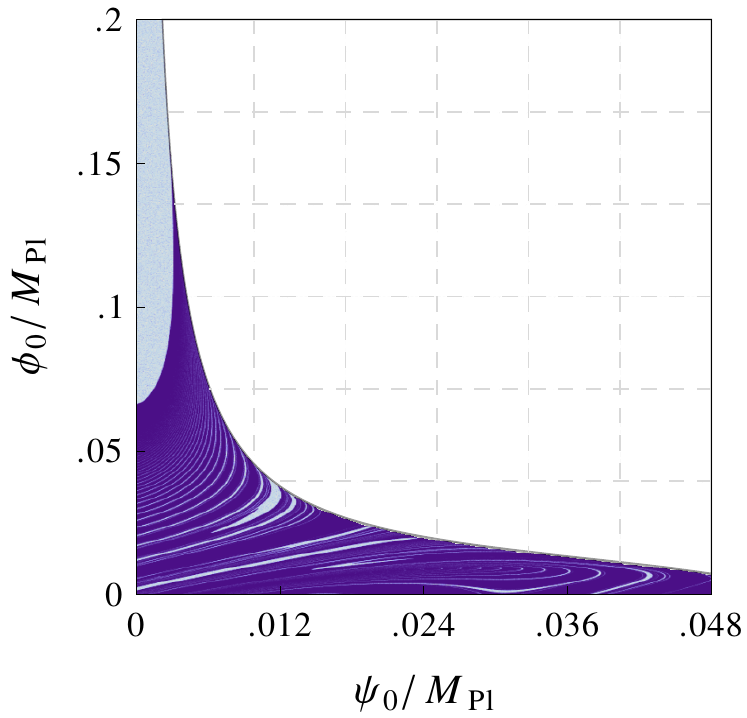}
      \includegraphics{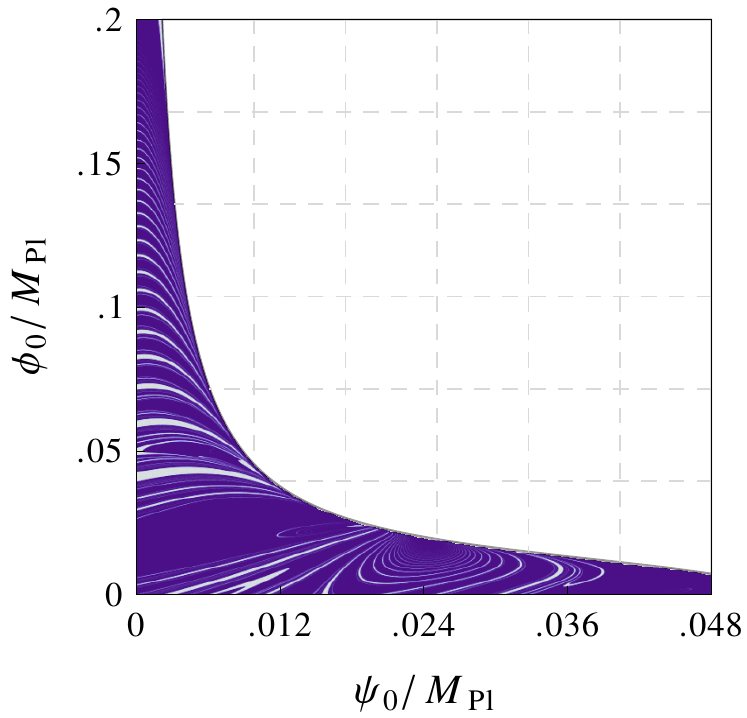}
      \caption{Two dimensional slicings of \sete{C} for $E=10^{-5} \, \mpl$, where the checked background has $V(\psi_0,\phi_0) > E^4$ and has not been sampled.  All velocities are of equal magnitude; the left panel has $\{\ivel{\psi},\ivel{\phi}\}>0$ and the right panel has $\{\ivel{\psi},\ivel{\phi}\}<0$.
      }
      \label{fig:slice1e-5}
    \end{figure}

   %

The fraction of successful points at any given energy $E$ is summarized in Table~\ref{table:ratio}, both including and excluding the ``false positives.''  The highest probability for success is at higher energies.  We should expect this since, given that the effective equation of state is not the same on all trajectories, orbits accumulate on the narrow inflationary attractor over time, leaving a larger flux of orbits through the attractor at lower energies.  In comparison to a sample drawn from $\sete{C}$, an identical sample from a slice $\mathcal C_{E'}$ with $E>E'$ will not place as much weight on trajectories inside the inflationary attractor and we expect to see fewer successfully inflating initial conditions on the lower energy surface.

Figures~\ref{fig:slice1e-2}~to~\ref{fig:slice1e-5} show  two dimensional slices of \sete{S} at different values of $E$.   We bin \sete{S} on a $1000 \times 1000$ grid: white regions are those with the highest number of successful points and the darkest regions have the fewest.  We accumulated 2.5 million successful points on each slice and see minor stochastic  variation in the number of points per bin.  Figures~\ref{fig:slice1e-2},~\ref{fig:slice1e-2_nofalsepos} ($E=10^{-2} \, \mpl$), and~\ref{fig:slice1e-5} ($E=10^{-5} \, \mpl$) show two-dimensional slices of  \sete{C} on which the  initial velocities have equal magnitude  $|\ivel{\phi}| = |\ivel{\psi}|$.  Looking at Figs~\ref{fig:slice1e-2}~through~\ref{fig:slice1e-5} we can see areas where \sete{S} and \sete{F} mix together, forming an intricate substructure similar to that seen in Refs~\cite{Clesse:2008pf,Clesse:2009ur}.  In Fig.~\ref{fig:slice1e-2_nofalsepos} we also see contiguous regions and thick bands which reliably inflate and survive the subtraction of the false positives from \sete{S}.  Qualitatively, \sete{S} exhibits considerable long range order when compared to Fig.~\ref{fig:clessefig1}.  At higher energy, contiguous regions in \sete{S} occupy a larger portion of \sete{C}, which can be seen clearly in Figs~\ref{fig:slice1e-2}~and~\ref{fig:slice1e-2_nofalsepos}.  The transition from Figs~\ref{fig:slice1e-2}~to~\ref{fig:slice1e-5} shows how the geometry of \sete{S} changes with $E$.

\begin{figure}
      \centering
      \includegraphics[width=.49\textwidth]{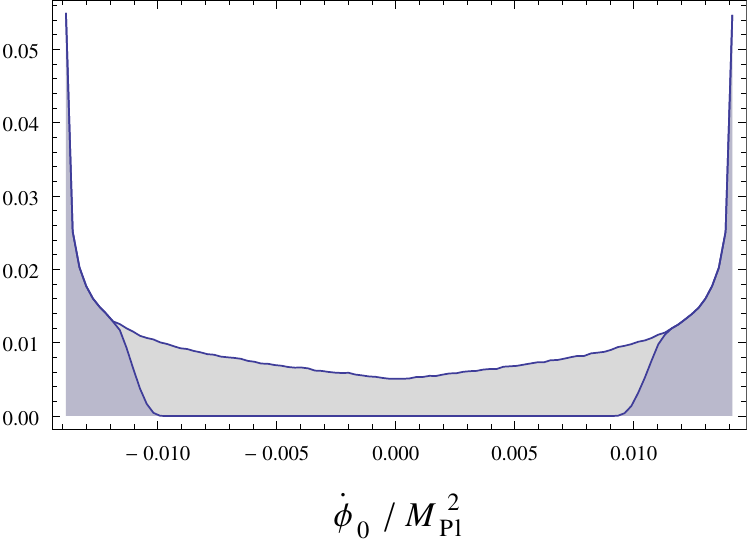}
      \includegraphics[width=.49\textwidth]{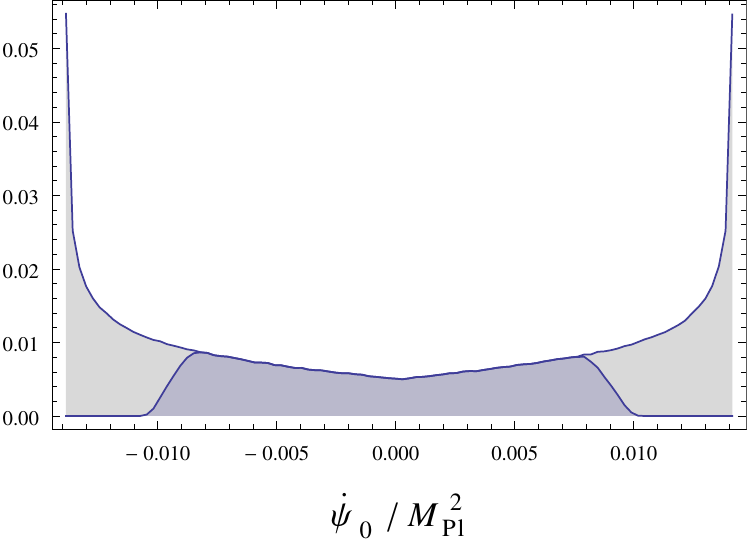}

      \includegraphics[width=.49\textwidth]{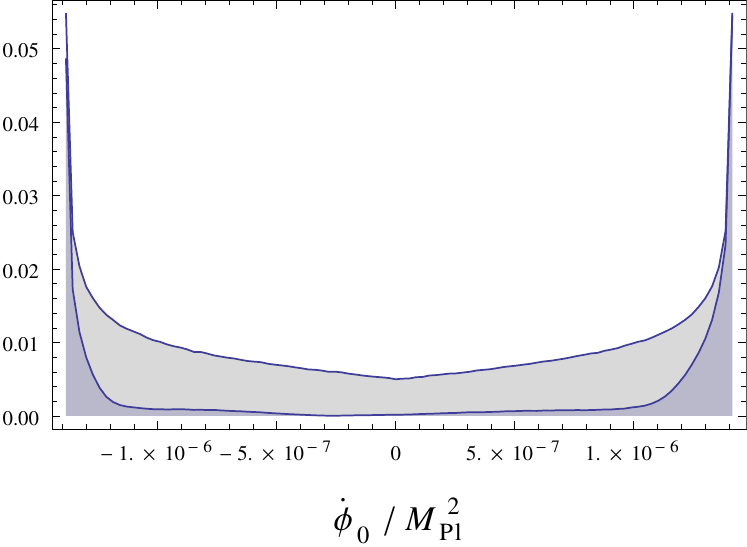}
      \includegraphics[width=.49\textwidth]{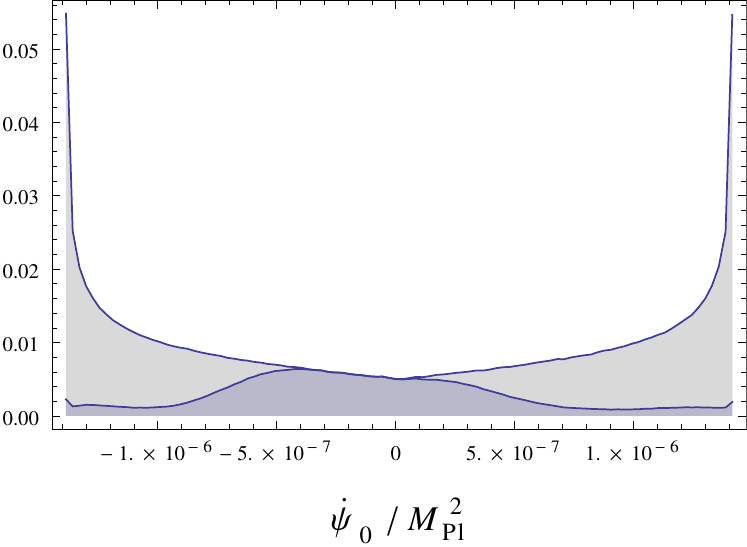}

      \includegraphics[width=.49\textwidth]{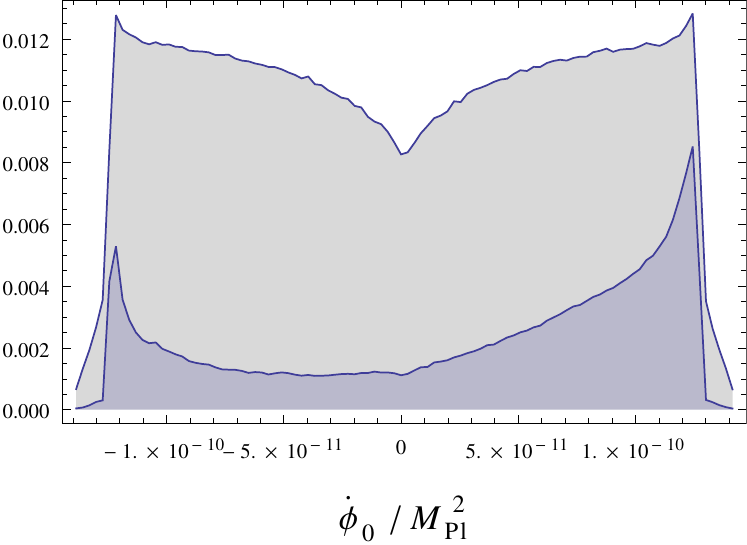}
      \includegraphics[width=.49\textwidth]{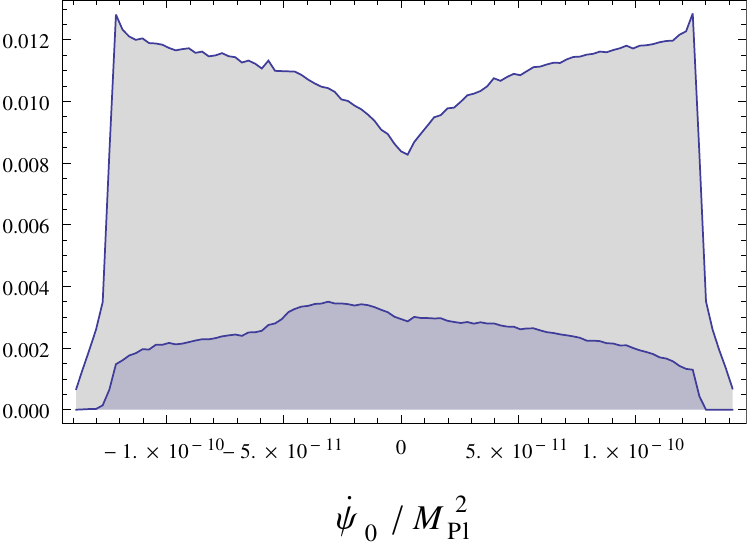}
      \caption{Histograms of $\ivel{\phi}$ (left column) and $\ivel{\psi}$ (right column) drawn from equal energy surfaces $\sete{C}$ with priors $P_\text{orig}$.  The rows have energy $10^{-1} \, \mpl$ (top), $10^{-3} \, \mpl$ (middle), and $10^{-5} \, \mpl$ (bottom).  The gray background is the total sample from $\sete{C}$ and the blue foreground is the successful subset \sete{S}. At $E=\mpl$ (not displayed), there are false positives only.}
      \label{fig:histogram}
    \end{figure}

Figures~\ref{fig:slice1e-2}~to~\ref{fig:slice1e-5} are projections of three dimensional regions and suppress information about the field velocities of the successful initial configurations. Intuitively, the points most likely to inflate (for given $\phi_0$ and $\psi_0$) would be those which had a large $|\ivel\phi|$ and small $|\ivel\psi|$.  These points are essentially ``launched'' up the inflationary valley, while the slope of the potential  focuses them toward smaller values of $\psi$.   To show this dependence on the initial values $\ivel\phi$ and  $\ivel\psi$, Fig.~\ref{fig:histogram} shows histograms of these values sampled from the whole of \sete{C}.  The fraction of sampled points in \sete{S} as a function of initial velocity confirms our intuition: most successful points have larger $|\ivel\phi|$ and smaller $|\ivel\psi|$.  Points for which $\ivel{\phi} \approx 0$ are particularly disfavored, further suggesting that the zero-velocity slice $\mathcal{Z}$ is unrepresentative of typical inflationary trajectories.  We can also see the impact of the ``false positives'' in these plots: these are more frequent at high energies and in the limiting case  $E=\mpl$ all na\"{i}vely-inflating initial conditions are false positives, since $\Delta$ encompasses the whole of $\mathcal{I}$ in this limit.

With sub-Planckian initial field values the kinetic energy dominates the potential energy for $E \gg \Lambda$.  Thus, even if a trajectory starts inside the inflationary valley, its velocity is such that it is unlikely to remain there.  For example, when $E=10^{-2} \, \mpl$ the slices of \sete{S} in Figs~\ref{fig:slice1e-2}~and~\ref{fig:slice1e-2_nofalsepos} show no special preference for points within the valley. In contrast, with  $E=10^{-5} \, \mpl \sim 10 \Lambda$ the valley is clearly distinguishable, as shown in Fig.~\ref{fig:slice1e-5}, but only when the  initial velocity of $\phi$ is directed ``uphill,'' i.e. with $\ivel{\phi} \ge 0$.  Conversely, each slice contains many successful points that lie outside the inflationary valley.

    \begin{figure}
     \centering
      \includegraphics{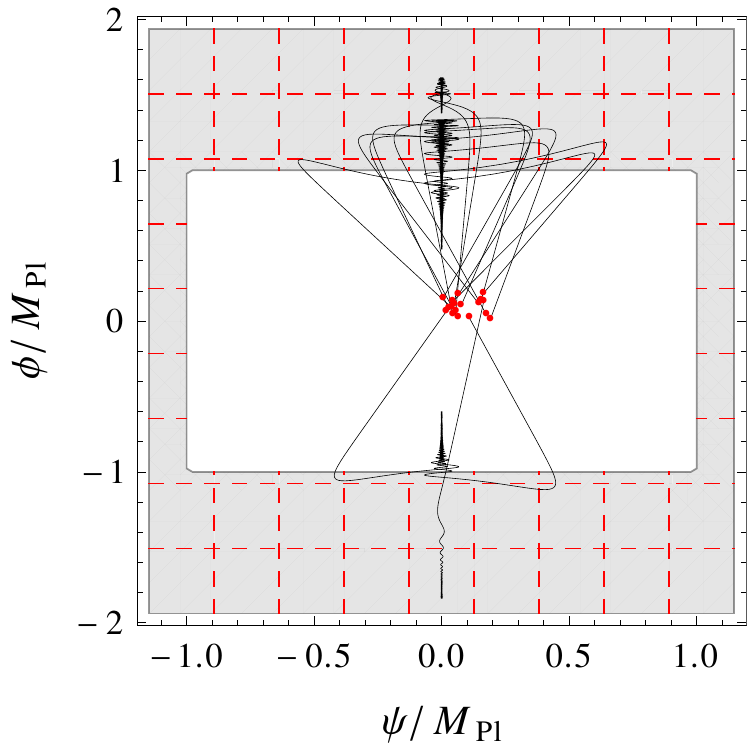}
      \includegraphics{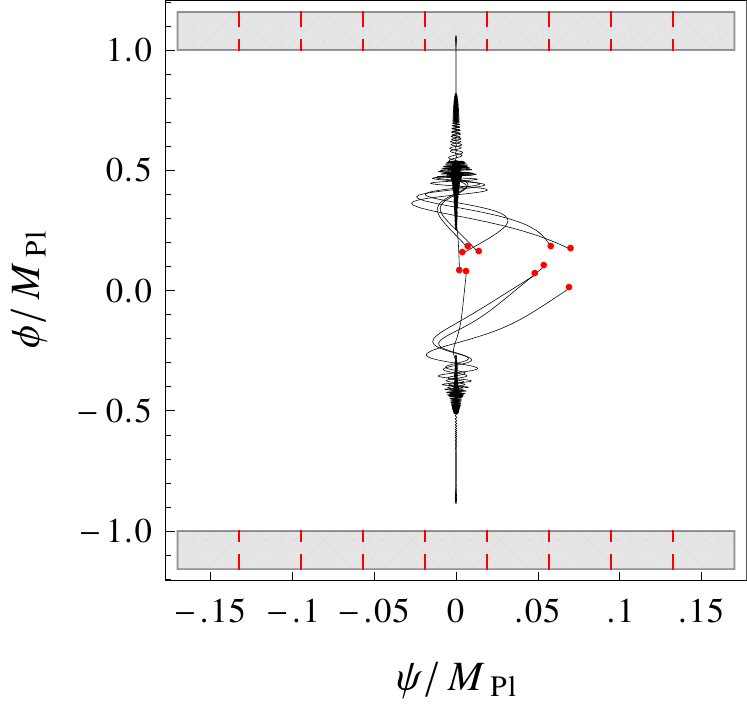}
      \caption{Successfully inflating trajectories projected onto the $\{\psi,\phi\}$ plane, with initial conditions $\{\psi_0,\phi_0\}$ marked in red.  The left panel is at energy $E=10^{-2} \, \mpl$ and the right panel is at energy $E=10^{-4} \, \mpl$.  The gray, checked region is where the magnitude of the field values exceeds $\mpl$.}
      \label{fig:trajectoryplot}
    \end{figure}

In Fig.~\ref{fig:trajectoryplot} we project specific representative solutions of Eqs~\eom onto the $\{\psi,\phi\}$ plane for initial conditions with energies $E=10^{-2} \, \mpl$ and $E=10^{-4} \, \mpl$.  Trajectories which unambiguously inflate show little topological mixing and are all reflected off of the maximum of the potential $V_\mathrm{max}=V|_{\psi=\phi}$ toward the inflationary valley.  For $E \gtrsim 10^{-3}\mpl$, most trajectories contain regions in which the field  values are super-Planckian. We do not exclude these trajectories, but we could easily add this as a separate requirement for a viable inflationary scenario, in which case almost no successful inflationary trajectories exist at these energies.

To quantify the sensitive dependence on initial conditions independently of our sampling procedure, we use the box-counting method to estimate the fractal dimension of both $\sete{S}$ and its boundary \sete{B}, including the ``false'' positives~\cite{theiler}.  We first cover each set $\sete{S}$ and $\sete{B}$ with progressively smaller four-dimensional boxes of size $\delta$, then count the number $N(\delta)$ of $\delta$-sized boxes in each covering. The box-counting dimension
\begin{equation}
    \label{eqn:boxdimn}
  d = \lim_{\delta \to 0} \frac{\log(N(\delta))}{\log(1/\delta)}
\end{equation}
is estimated by the slope of the line fitted to the linear portion of the curve $\log(N)$ as a function of $\log(1/\delta)$.  To compute the dimension of \sete{B} we count boxes that contain elements of both \sete{S} and \sete{F}.
  \begin{figure}
    \centering
      \includegraphics[width=.47\textwidth]{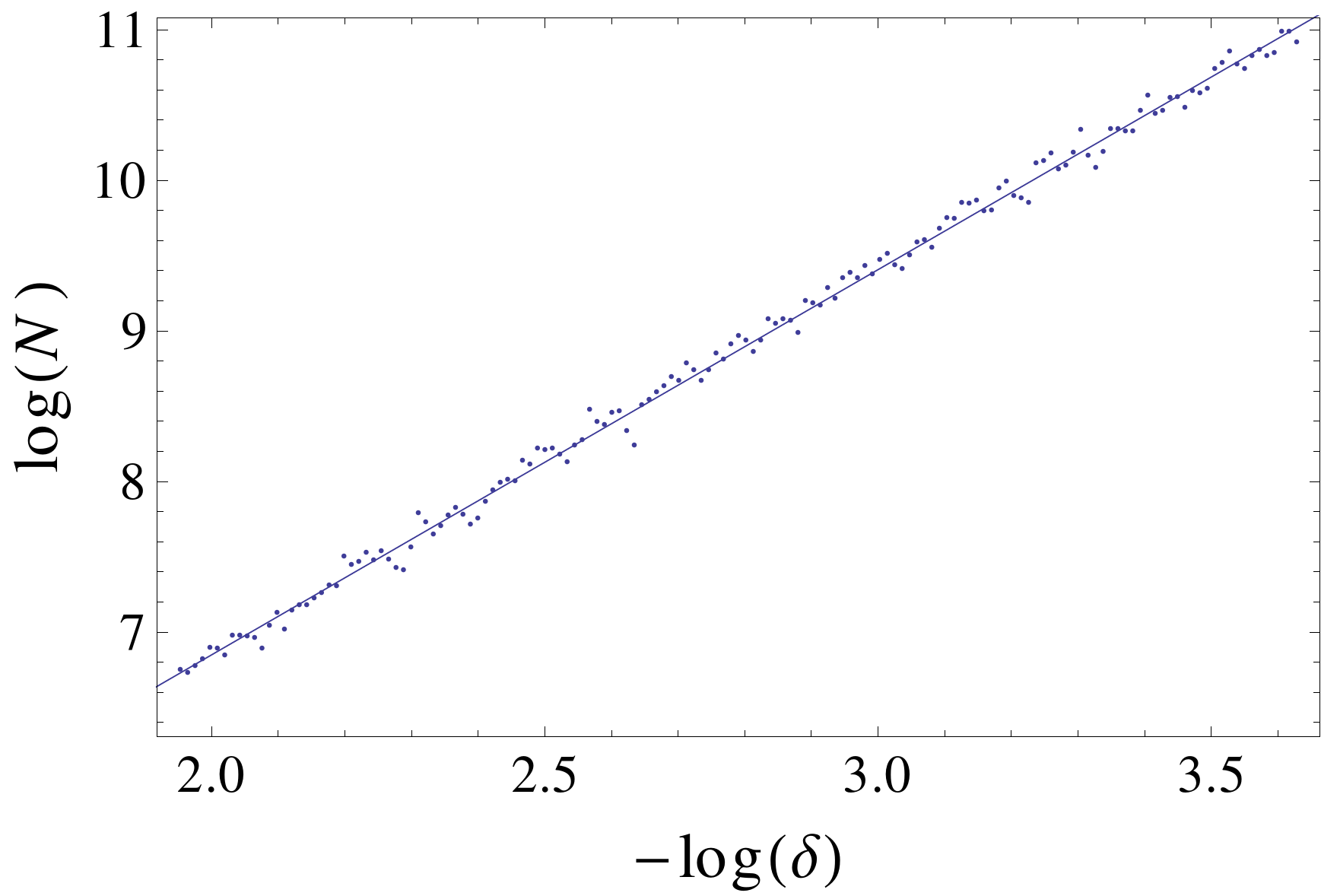}
      \includegraphics[width=.47\textwidth]{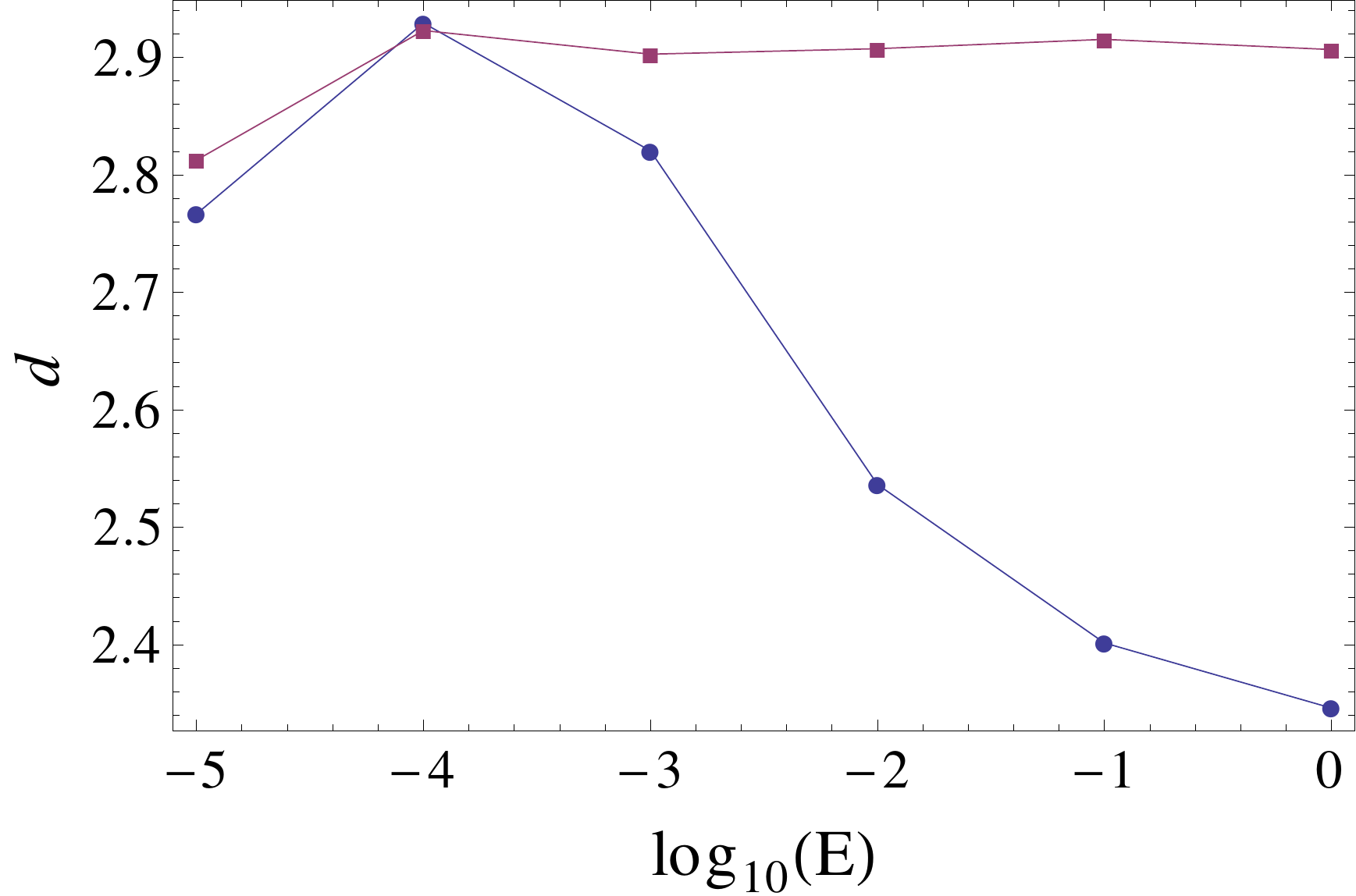}
     \caption{The left panel shows the box counting of \sete{B} for $E=10^{-2} \, \mpl$.  The slope of the best-fit line $d=2.558$ is the box counting dimension.  The right panel is the box counting dimension $d$ versus the energy $E$ for sets \sete{S} and \sete{B} at the energies in Eq.~\eqref{eqn:chosenenergies}.  The red line with boxes indicates \sete{S} and the blue line with circles is \sete{B}.}
    \label{fig:fractaldimn}
  \end{figure}
  Figure~\ref{fig:fractaldimn} shows both a typical fit (for \sete{B} with $E=10^{-2} \, \mpl$)  and the computed values of $d$ for   \sete{S} and \sete{B}.  The result is sensitive to the detailed fitting procedure, which we trained by testing the algorithm on sets with known dimension, such as Cantor dust.  Furthermore, the estimate for $d$ depends both on the non-trivial distribution of $\sete{S}$ over $\sete{C}$ and the resolution of sampled points, which is a function of the initial energy and sampling prior.  The reported values of $d$ should be interpreted as an upper bound to the more fundamental Hausdorff dimension~\cite{theiler} that improves with increasing $E$, where the set \sete{S} has higher long-range order.

The regions considered here are multifractal, in that the dimension of   \sete{S} will be a function of both position in $\mathcal{I}$ and the overall scale. The first is easy to see: each surface \sete{C} contains regions in which essentially all points inflate (in these regions $d \approx 3$) and regions that are approximately isolated points (with $d<3$). Consequently, the computed value of $d$ is effectively a weighted average of at least two different regions, which explains why the dimension of \sete{S} is close to 3 but still measurably non-integer.  Secondly, at very small scales \sete{S} must consist of smooth contiguous regions and on these scales we expect $d\rightarrow3$.  These regions exist in spite of the chaotic dynamics due to the dissipative terms in Eqs~\eqref{eqn:hybrideom}--\eqref{eqn:hybridv} and put a lower limit on the mixing scale.
  
%

\subsection{The role of the prior}
\label{ssect:prior}

It is well-known \cite{Gibbons1987,Hawking:1987bi,Felder:1999pv,Gibbons:2006pa,Freivogel:2011eg,Schiffrin:2012zf,Downes:2012xb} that probability measures on different hypersurfaces result in different conclusions regarding the likelihood of inflation; we explore here how this relates to the choice of sampling prior.  Although surfaces with different energies (as well as different initial conditions surfaces, such as a slice of constant comoving time) are homeomorphic to \sete{C} and, by definition, have the same topology, the prior on \sete{C} is not a topological property and is not preserved under either homeomorphism or a change of variables.\footnote{A homeomorphism is provided by time-translation along the integral curves of the equations of motion.  Note that $\mathcal{Z}$, being of a lower dimension, is not homeomorphic to \sete{C}.}  Each initial condition surface then has a different prior and different likelihood for inflation, even given the same sampling technique.  

    \begin{table}
      \centering
      \begin{tabular}{| c |  c  c |  c  c |  c c | c c | }
        \hline
        --- & \multicolumn{2}{c |}{$P_\mathrm{orig}$} & \multicolumn{2}{c |}{$P_\mathrm{square}$} & \multicolumn{2}{c |}{$P_{\dot{\phi}}$} & \multicolumn{2}{c |}{$P_{\dot{\psi}}$} \\
        \hline
        $E$ [$\mpl$] & $f_\mathrm{true}$ & $f_\mathrm{total}$ & $f_\mathrm{true}$ & $f_\mathrm{total}$& $f_\mathrm{true}$ & $f_\mathrm{total}$& $f_\mathrm{true}$ & $f_\mathrm{total}$  \\
        \hline
        $10^{0}$             & 0.000  & 0.498 &  0.000& 0.497& 0.000& 0.704& 0.000& 0.294  \\
        $10^{-1}$            & 0.447  & 0.479 &  0.426& 0.471& 0.651& 0.681& 0.243& 0.276\\
        $10^{-2}$            & 0.389  & 0.434 &  0.347& 0.408& 0.582& 0.624& 0.196& 0.242 \\
        $10^{-3}$            & 0.300  & 0.331 &  0.245& 0.285& 0.438& 0.469& 0.162& 0.194 \\
        $10^{-4}$            & 0.205  & 0.211 &  0.188& 0.195& 0.247& 0.253& 0.163& 0.169 \\
        $10^{-5}$            & 0.225  & 0.225 &  0.215& 0.215& 0.265& 0.265& 0.183& 0.183 \\
        \hline
      \end{tabular}
      \caption{Fraction $f$ of sampled points from \sete{C} that inflate --- both excluding (true) and including (total) false positives. The  sampling techniques ($P_\mathrm{orig}$, $P_\mathrm{square}$, $P_{\dot{\phi}}$, and $P_{\dot{\psi}}$) are explained in the text.}
    \label{table:prior}
  \end{table}

  In Table~\ref{table:prior} we compare different uninformative priors, defined implicitly through four sampling algorithms, on surfaces \sete{C} at the energies in Eq.~\eqref{eqn:chosenenergies}.  Since the kinetic energy is initially dominant for the energies and ranges we consider, we leave the selection method for $\{\psi_0, \phi_0\}$ the same as in Section~\ref{ssect:method}, but vary the way we set the velocities $v \in \{\ivel{\psi}, \ivel{\phi} \}$.  The original prior $P_\mathrm{orig}$ draws one velocity $v_1$ from a uniform distribution, bounded by $\pm \sqrt{2(E^4 -V_0)}$, and then sets $v_2$ by the energy constraint~\eqref{eqn:energyeqn}.  All signs are chosen randomly and this procedure is alternated on subsequent choices of points to obtain a symmetric distribution in the velocities.  The second prior $P_\mathrm{square}$ is similar, except we draw the \emph{square} of the velocity $v_1^2$ from a uniform distribution bounded below by zero and above by $2(E^4-V_0)$, with the sign of $v_1$ chosen randomly.  Again, we alternate this to obtain a symmetric distribution.  With this modest change in prior, the calculated fraction $f_\mathrm{true}$ of \sete{C} that inflates (excluding false positives) differs by only a few percent, with $P_\mathrm{square}$ giving a slightly lower fraction at each energy.  The fraction $f_\mathrm{true}$ again decreases with decreasing $E$.

We compare this to two priors $P_{\dot \phi}$ and $P_{\dot \psi}$ that are asymmetric in the velocities.  For $P_{\dot \phi}$ we always draw $\ivel{\psi}$ from a uniform distribution bounded by $\pm \sqrt{2(E^4 -V_0)}$ and always set $\ivel{\phi}$ by the energy constraint.  For $P_{\dot \psi}$ we do the opposite: draw $\ivel \phi$ and set $\ivel \psi$.  This gives a uniform distribution in the sampled velocity $v_1$, but a high-tail distribution similar to Fig.~\ref{fig:histogram} in the velocity $v_2$.
The prior $P_{\dot \phi}$ focuses more of the sample around $\ivel{\psi} \approx 0$, the area identified as being most-likely to inflate, and $P_{\dot \psi}$ gives more points around $\ivel{\phi} \approx 0$, the area least-likely to inflate.  At $E=\mpl$ the difference in $f_\mathrm{total}$ (including false positives) between the asymmetric priors is as much as 41.0 percentage points.  The differences  decrease with decreasing $E$, indicating that later-time hypersurfaces become progressively independent of the prior.
However, Table~\ref{table:prior} demonstrates how any measure of $f$ is prior-dependent, especially with respect to the implicit dependence of the prior on the initial energy.


\section{Conclusion}
\label{sect:conclusion}

We have considered the initial conditions problem for multifield inflation, quantifying the likelihood of inflation by sampling an initial conditions surface, evolving the points numerically, and  dividing them into successfully and unsuccessfully inflating sets.  We draw initial conditions from an equal energy slice of phase space, denoted \sete{C}, the maximum energy at which the underlying theory is assumed to be an accurate description of the primordial universe.  Since FLRW universes have a monotonic energy density, sampling initial conditions from \sete{C} ensures that we count only unique solutions to the equations of motion.  A sample of points  from \sete{C} is thus a well-defined sample of homogeneous universes.  Typically, we cannot predict the flux of orbits through \sete{C} and must choose a prior, accordingly.   We considered four different uninformative priors on \sete{C} and showed
that the likelihood of inflation varies by as much as a factor of roughly two between candidate priors.  However, one can imagine scenarios where the prior dependence was much more dramatic.

After specializing to hybrid inflation we examined the topology of the set of successful points \sete{S}, which is independent of continuous deformations to the prior.  We confirm that both \sete{S} and the boundary between the successful and unsuccessful points is fractal for all sampled energies.  The structure of \sete{S}, as seen in Figs~\ref{fig:slice1e-2}~to~\ref{fig:slice1e-5}, is qualitatively smoother than when initial conditions are chosen from the zero-velocity slice shown in Fig.~\ref{fig:clessefig1}.  Further, since the equations of motion~\eom are dissipative, there must be a small-scale cutoff to any fractal structure.  However, quantum fluctuations put a fundamental lower limit on the homogeneity of the early universe: if  $\sete{S}$ has structure below this scale, the assumption of homogeneity is not self-consistent. Fluctuations are larger at higher energies and above some critical energy $E$ the number of viable, homogeneous scenarios is vanishingly small, even though the na\"{\i}ve  counting statistic  suggests that a nontrivial fraction of the initial conditions space is inflationary.

Our specific calculations are performed for the  hybrid potential~\eqref{eqn:hybridv}, but our underlying goal is to develop tools that can be applied to the initial conditions problem associated with generic multifield scenarios.  Recent progress has been made by studying both random multifield models \cite{Aazami:2005jf,Tye:2008ef,Frazer:2011br,Marsh:2011aa,Battefeld:2012qx,McAllister:2012am} and inflection point models \cite{Itzhaki:2007nk,Allahverdi:2008bt,Itzhaki:2008hs,Spalinski:2009rq,Downes:2012xb,McAllister:2012am}.  These approaches yield contrasting conclusions regarding the distribution of inflationary trajectories; applying the methods developed here   to  these models will be an interesting extension of this work.

This analysis  assumes that the universe is initially spatially flat and homogeneous, but even if     inflation begins without tuning in the homogeneous limit there is no guarantee that this result will survive  the addition of pre-inflationary inhomogeneities.  Inhomogeneous pre-inflationary configurations were  examined by Goldwirth and Piran \cite{Goldwirth:1989pr,Goldwirth:1990pm,Goldwirth:1991rj}, who showed that single-field chaotic inflation  and new inflation \cite{Linde:1981mu,Albrecht:1982wi} remain robust in the presence of nontrivial inhomogeneity, provided that the initial field value is approximately correlated over several Hubble radii. We plan to examine this question for multifield inflation in future work.

\acknowledgments

The authors acknowledge the contribution of the NeSI high-performance computing facilities and the staff at the Centre for eResearch at the University of Auckland. New Zealand's national facilities are provided by the New Zealand eScience Infrastructure (NeSI) and funded jointly by NeSI's collaborator institutions and through the Ministry of Science \& Innovation's Research Infrastructure programme [{\url{http://www.nesi.org.nz}}].   We thank the Yukawa Institute for Theoretical Physics at Kyoto University, where a draft of this work was presented at workshop YITP-T-12-03, and Grigor Aslanyan, S\'{e}bastien Clesse, and Christophe Ringeval for helpful comments on the manuscript.  We also thank the Center for Applied Scientific Computing at LLNL for making the \textsc{Sundials} package freely available.

\bibliographystyle{JHEP}

\bibliography{references}

\end{document}